\documentclass{article}

\usepackage{arxiv}

\usepackage[utf8]{inputenc} 
\usepackage[T1]{fontenc}    
\usepackage{hyperref}       
\usepackage{url}            
\usepackage{booktabs}       
\usepackage{amsfonts}       
\usepackage{nicefrac}       
\usepackage{microtype}      
\usepackage{lipsum}		
\usepackage{graphicx}
\usepackage{natbib}
\usepackage{doi}
\usepackage{textcomp, gensymb}
\usepackage{amsmath}
\usepackage{amssymb}
\usepackage{latexsym}
\usepackage{mathtools}
\usepackage{breqn}

\title{Advanced ensemble modeling method for space object state prediction accounting for uncertainty in atmospheric density}

\author{Smriti Nandan Paul \\
	Department of Mechanical and Aerospace Engineering\\
	West Virginia University\\
	Morgantown, WV 26505 \\
	\texttt{smritinandan.paul@mail.wvu.edu} \\
	\And
	{Richard J. Licata} \\
	Department of Mechanical and Aerospace Engineering\\
    West Virginia University\\
	Morgantown, WV 26505 \\
	\texttt{rjlicata@mix.wvu.edu} \\	
	\And
	{Piyush M. Mehta} \\
	Department of Mechanical and Aerospace Engineering\\
    West Virginia University\\
	Morgantown, WV 26505 \\
	\texttt{piyush.mehta@mail.wvu.edu} \\		
}

\renewcommand{\shorttitle}{}

\begin{document}
\maketitle

\begin{abstract}
	For objects in the low Earth orbit region, uncertainty in atmospheric density estimation is an important source of orbit prediction error, which is critical for space situational awareness activities such as the satellite conjunction analysis. This paper investigates the evolution of orbit error distribution in the presence of atmospheric density uncertainties, which are modeled using probabilistic machine learning techniques. The recently proposed ``HASDM-ML,'' ``CHAMP-ML,'' and ``MSIS-UQ'' machine learning models for density estimation \citep{Licata2022, msisuq_arxiv} are used in this work. The investigation is convoluted because of the spatial and temporal correlation of the atmospheric density values. We develop several Monte Carlo methods, each capturing a different spatiotemporal density correlation, to study the effects of density uncertainty on orbit uncertainty propagation. However, Monte Carlo analysis is computationally expensive, so a faster method based on the Kalman filtering technique for orbit uncertainty propagation is also explored. It is difficult to translate the uncertainty in atmospheric density to the uncertainty in orbital states under a standard extended Kalman filter or unscented Kalman filter framework. This work uses the so-called ``consider covariance sigma point (CCSP)'' filter that can account for the density uncertainties during orbit propagation. As a test-bed for validation purposes, a comparison between CCSP and Monte Carlo methods of orbit uncertainty propagation is carried out. Finally, using the HASDM-ML, CHAMP-ML, and MSIS-UQ density models, we propose an ensemble approach for orbit uncertainty quantification for four different space weather conditions. 
\end{abstract}


\section{Introduction}
\label{sec1}
Ambitious satellite mega-constellation projects (e.g., SpaceX's Starlink \citep{McDowell_2020}, OneWeb satellite constellation \citep{Henri2020}) and affordable access to space have led to an exponential growth of objects in the low Earth orbit (LEO) region in recent decades. There are no signs of a reversal of this trend, as tens of thousands of satellites are tentatively planned for launch in the near future \citep{Boley2021}. This proliferation increases the risk of collisions between active space assets and space debris or between debris and debris, threatening the sustainability of this commercially and scientifically critical near-Earth region. We need better space situational awareness (SSA) measures to address this sustainability challenge. A particularly important aspect of SSA is the conjunction assessment that involves computing the probability of collision between two space objects, which is critical for operational decisions such as the firing of thrusters or differential drag application \citep{finley2013techniques} for orbit modification. Reliable estimates of the uncertainties in the orbital state of a space object are required for determining the probability of collision. This paper thus focuses on the quantification and propagation of orbital state uncertainties.

The primary sources of errors in orbit prediction problems are - (a) initial state uncertainties, (b) dynamical uncertainties, and (c) uncertainties from sensor measurements of space objects. In the proposed study, we focus on the effects of uncertainties in atmospheric density, which is the largest source of dynamical uncertainties for objects in the LEO region. The uncertainty in atmospheric density manifests itself through the atmospheric drag force, which is given as:

\begin{equation}\label{drag_accn}
\vec{a}_{D} = -\frac{1}{2}\rho \frac{C_D A_{proj}}{m}v_{rel}\vec{v}_{rel}
\end{equation}
where $\rho$ is the atmospheric density, $C_D$ is the drag coefficient, $A_{proj}$ is the projected area of the satellite perpendicular to the flow direction, $\vec{v}_{rel}$ is the velocity of the satellite relative to the atmosphere, and $v_{rel}$ is the magnitude of $\vec{v}_{rel}$. The satellite mass is usually known from the operators and constant for a non-maneuvering satellite, but all other parameters can have uncertainties \citep{VALLADO2014141}. Except for the atmospheric neutral mass density, we will not address the uncertainties in other drag parameters in this paper. 

Atmospheric density and its associated uncertainty have a complex dependency upon the selected atmospheric model, solar irradiance in the extreme ultraviolet (EUV) and far ultraviolet (FUV) spectral ranges, geomagnetic indices, location, epoch, and various other factors, which make it a challenging problem to estimate the density. The difficulty is particularly amplified during large geomagnetic storms \citep{bruinsma2021} and can even lead to the loss of satellites \citep{2022SW003074}. Several thermospheric mass density models have been proposed over the years. These density models can either be categorized as physical models (those that solve fluid equations) or empirical models (those that represent average behavior of atmospheric observations in a parameterized mathematical formulation \citep{EMMERT2015773}). A detailed review of existing models can be found in \cite{HE201831, EMMERT2015773}. However, most density models in existence, including the operational High Accuracy Satellite Drag Model (HASDM) system \citep{STORZ20052497} used by the United States Space Force (USSF), do not provide an estimate of the uncertainty in their predictions.

Little attention has been paid in the literature to the direct quantification of the uncertainty in atmospheric density models. Using Gaussian Processes (GPs), \cite{GAO2020273} propose a framework for uncertainty quantification of NRLMSISE-00 (Naval Research Laboratory Mass Spectrometer and Incoherent Scatter Radar Extended) and JB2008 (Jacchia-Bowman) neutral mass density models. \cite{refId0} use statistical data binning techniques combined with least square fitting approaches to provide an uncertainty quantification model for the DTM2020 (Drag Temperature Model) thermosphere density model. \cite{licatadendropout} leverage the Monte Carlo dropout technique, a Bayesian approximation of the Gaussian Process, to develop a probabilistic density model utilizing the Space Environment Technologies (SET) High Accuracy Satellite Drag Model (HASDM) density database. More recently, in two different works - the first by \cite{Licata2022} and the second by \cite{msisuq_arxiv} - the authors have developed three machine learning models that directly predict mean and standard deviation (as opposed to an ``ensemble-like'' approach followed in the computationally expensive Monte Carlo dropout technique). The first model, which we will refer to as the HASDM-ML-DP (\textit{ML}: machine learning; \textit{DP}: direct probability prediction), is based on the SET HASDM database; the second model, which we will refer to as the CHAMP-ML-DP, is based on the accelerometer data collected by the Challenging Minisatellite Payload (CHAMP) satellite; the third model, which we will refer to as the MSIS-UQ-DP, is based on combined data from the CHAMP, Gravity Recovery and Climate Experiment (GRACE), Swarm A, and Swarm B satellites. The methodology we develop in this paper will use the HASDM-ML-DP, CHAMP-ML-DP, and MSIS-UQ-DP thermospheric density models \citep{Licata2022, msisuq_arxiv}, which are detailed in later sections of this paper.

Several authors have investigated the effects of atmospheric density uncertainty on the orbital state uncertainties or derived quantities such as the probability of collision. \cite{wilkins2000characterizing} use a first-order Gauss-Markov process to model atmospheric density uncertainty in an Extended Kalman Filter (EKF) framework for orbit determination. \cite{sagnieres2017uncertainty} present an Ornstein-Uhlenbeck process-based framework that uses the intrinsic difference between various atmospheric density models to characterize the uncertainty in atmospheric density and subsequently study its effect on orbit prediction. \cite{EMMERT2017147}  develop analytic expressions for in-track position errors due to EUV forecast errors modeled using the Brownian motion process. \cite{BussyVirat18} investigate the effects of uncertainty in \textit{F\textsubscript{10.7}} and \textit{Ap} space weather drivers on the probability of collision. Using Proper Orthogonal Decomposition (POD), \cite{Gondelach22} derive a dynamic reduced-order density model, which is then used in a Kalman filtering framework for uncertainty propagation. The authors use the estimated uncertainties for the calculation of the probability of collision. However, none of these works use an atmospheric density model that explicitly provides the uncertainty in its estimate.

The goal of the present paper is to develop an ensemble methodology that combines the epistemic uncertainties predicted by the HASDM-ML-DP, CHAMP-ML-DP, and MSIS-UQ-DP density models to characterize the uncertainty in the orbital states of a space object. In section 2, we describe the data and approach used to develop the three stochastic density models. In section 3, we demonstrate how a standard high-cadence Monte Carlo technique to orbit uncertainty propagation is insufficient for capturing the spatiotemporal density correlation. We then provide several Monte Carlo orbit propagation algorithms that not only account for the effect of density uncertainty but also preserve spatiotemporal correlation for the density. In section 4, we provide details about the so-called consider covariance sigma point (CCSP) filter that is used for orbit uncertainty propagation while capturing the effects of atmospheric density uncertainty. We present details of a framework that uses ensemble modeling for predicting the orbit state probability density function (PDF) in section 5. In section 6, test cases for the ensemble approach for orbit uncertainty characterization are presented for a variety of space weather conditions. In the same section, we also provide a comparison of the CCSP and Monte Carlo methods for validation purposes. Finally, in section 7, we summarize the paper and draw important conclusions.

\section{Stochastic Density Models}
\subsection{HASDM-ML-DP Density Model}
The HASDM is a proprietary thermospheric density framework maintained by the SET and currently in use for the USSF operations. Using the so-called Dynamic Calibration of the Atmosphere (DCA) algorithm, the HASDM framework estimates 13 global density correction parameters to provide near real-time corrections to the base JB2008 thermospheric density model. The computation of the global density correction field relies on the observed drag effects on a large number of calibration satellites in the LEO region. For a thorough discussion on the HASDM framework, see \cite{STORZ20052497}.

Recently, as part of an open-access initiative for scientific studies, SET has publicly made available the SET HASDM density database \citep{sethasdm2021} consisting of data from January 1, 2000 through December 31, 2019. The publicly available data has a cadence of three hours and a resolution of $15\degree$ longitude, $10\degree$ latitude, and 25 km altitude ranging between 175–825 km. The database consists of 58,440 samples. \cite{Licata2022} use this database for developing a deep neural network (DNN)-based framework that directly predicts the mean and standard deviation of parameters of interest, which are subsequently processed through an inverse function to obtain estimates of the mean and standard deviation of the atmospheric density. The inputs for their DNN, motivated by the drivers for the JB2008 density model, consist of eight solar indices/proxies, 16 geomagnetic indices (time history for \textit{ap} and \textit{Dst}), and four temporal parameters (for more details on the inputs, see \cite{Licata2022}). 

The SET HASDM database has 12,312 outputs (combination of 24 longitude values, 19 latitude values, and 27 altitude values) at each epoch, which would make any regression efforts a computationally challenging task. To facilitate computational feasibility, \cite{Licata2022} use principal component analysis (PCA) to obtain a reduced-order model (ROM) \citep{MehtaL2017SW001642, MLS2018}. The authors use PCA to perform the following decomposition:

\begin{subequations}
\begin{equation}
    \textbf{x}(\textbf{s}, t) = \Bar{\textbf{x}}(\textbf{s}) + \Tilde{\textbf{x}}(\textbf{s}, t)
\end{equation}
\begin{equation}
    \Tilde{\textbf{x}}(\textbf{s}, t) = \sum_{j=1}^{10} [\alpha_j(t) U_j(\textbf{s})]
\end{equation}
\end{subequations}
where $\textbf{x}(\textbf{s}, t)$ is the log-transformed HASDM density, $\Bar{\textbf{x}}(\textbf{s})$ is the mean dependent only on the spatial coordinates, $\alpha_j(t)$ are the temporal PCA coefficients, and $U_j(\textbf{s})$ are the orthogonal modes or basis functions. In the HASDM-ML-DP, the authors predict the mean and standard deviation of the 10 PCA coefficients, i.e., the output dimension of the ML model is 20. The predicted coefficients are multiplied by matrix $U$ (formed from the orthogonal modes of variation), followed by the addition of the spatial mean to decode back to log density. The authors carry out a Monte Carlo simulation, where multiple sets of PCA coefficients are sampled from their distribution, followed by the mentioned multiplication/addition operation to obtain a stochastic estimate of the log density. For a detailed description of the computation of the matrix $U$, refer to \cite{QualHASDM}.

\subsection{CHAMP-ML-DP Density Model}
CHAMP was a German satellite launched on July 15, 2000 \citep{REIGBER2002129} and remained in orbit for a decade before re-entering the Earth's atmosphere in 2010. Its orbit was near-circular, near-polar ($i \sim 87 \degree$), and an initial altitude of 460 km \citep{REIGBER2002129} was chosen. The STAR accelerometer on board the CHAMP satellite measured the resultant non-gravitational forces experienced by the satellite. By modeling out the effects of atmospheric lift, solar radiation pressure, albedo, and infrared radiation pressure, \cite{suttondiss} derives estimates of acceleration due to atmospheric drag. Thereafter, Sutton uses a drag coefficient model ($C_{D_{Sutton}}$) and a satellite geometry (representative of the satellite cross-sectional area, $A_{sutton}$) to obtain an estimate of the neutral thermospheric density ($\rho_{Sutton}$). However, Sutton's work uses a simplified model of the satellite drag coefficient. Using a higher fidelity satellite geometry ($A_{Mehta}$) and an improved drag coefficient model ($C_{D_{Mehta}}$), \cite{2016SW001562} scales Sutton's density estimates to obtain a new set of density estimates for the CHAMP satellite as:

\begin{equation}
    \rho_{Mehta} = \frac{C_{D_{Sutton}}A_{sutton}}{C_{D_{Mehta}}A_{Mehta}}\rho_{Sutton}
\end{equation}

The CHAMP database provided by \cite{2016SW001562} spans from January 1, 2002 through February 22, 2010. The data has a cadence of 10 seconds, making it a total of more than 25 million samples. The CHAMP-ML-DP \citep{Licata2022} is a DNN-based regression model that utilizes the density data provided by \cite{2016SW001562}. The input data for the CHAMP-ML-DP consists of eight solar indices/proxies, three geomagnetic indices, and eight spatiotemporal parameters, and the output data consists of the mean and standard deviation of the atmospheric density. More details on the input and the training of the model can be found in \cite{Licata2022}. The main distinction between the CHAMP-ML-DP and the HASDM-ML-DP is that the CHAMP-ML-DP is based on local measurements, whereas the HASDM-ML-DP is based on global measurements.

\subsection{MSIS-UQ-DP Density Model}
The Naval Research Laboratory updated the empirical NRLMSISE-00 density model to release a new version called NRLMSIS 2.0 \citep{NRLMSIS2} in 2021. Changes are incorporated in the fundamental formulations, and substantial additional measurements are included to make NRLMSIS 2.0 more accurate than the original version. The density estimates from  NRLMSIS 2.0 depend upon the exospheric temperature. Recently, \cite{msisuq_arxiv} have developed a feed-forward deep neural network model, MSIS-UQ-DP, that performs uncertainty quantification for the exospheric temperature, which in turn is fed into the NRLMSIS 2.0 model to obtain uncertainty estimates for the density. To develop the MSIS-UQ-DP model, the authors use a database of 81 million exospheric temperatures. These exospheric temperatures are computed using a binary search method \citep{intercal,extemplar,comparison} such that the densities from NRLMSIS 2.0 match density estimates of the CHAMP, GRACE-A, Swarm A, and Swarm B satellites. CHAMP and GRACE-A density estimates are obtained from accelerometer measurements, whereas the Swarm density estimates are obtained from orbit determination using the onboard Global Positioning System (GPS) receivers \citep{extemplar}. The input data for the MSIS-UQ-DP model consists of 21 space weather, spatial, and temporal parameters. The MSIS-UQ-DP model provides a well-calibrated uncertainty estimate and has a superior mean absolute error performance compared to the NRLMSIS 2.0 and HASDM density models. For more details on the MSIS-UQ-DP model, please refer to \cite{msisuq_arxiv}.

\section{Monte Carlo Methods for Orbit Uncertainty Propagation}

\subsection{`Traditional' Monte Carlo Method With a High Sampling Frequency}
\begin{figure*}[hbt!]
\centering
\includegraphics[width=.8\textwidth]{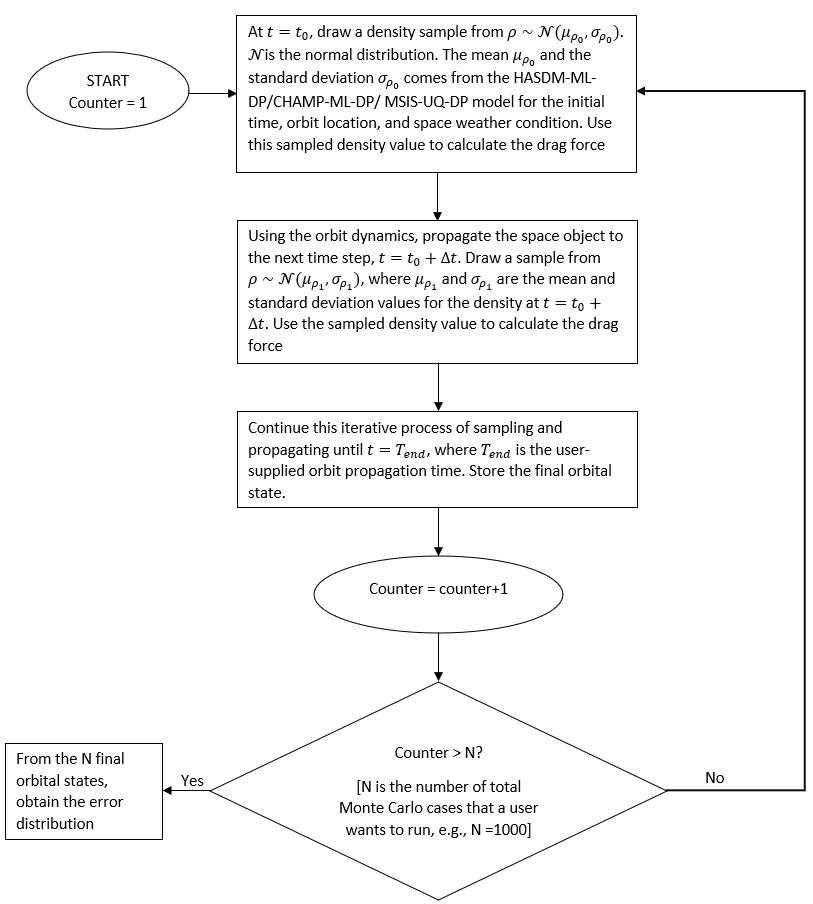}
\caption{`Traditional' Monte Carlo Method for orbit uncertainty propagation.}
\label{trad_mc}
\end{figure*}

Orbital parameters and measurements have uncertainties, which are expressed using PDFs derived from some mathematical model or heuristics. In a Monte Carlo method, one repeatedly samples from the PDF representing the uncertainty and carries out orbit propagation to construct a population of objects, from which statistical information about their states can be obtained. The schematic of a `traditional' Monte Carlo approach to orbit uncertainty propagation in the presence of atmospheric density uncertainties is shown in Fig. \ref{trad_mc}. In the traditional approach, the sampling frequency, which is defined as the inverse of $\Delta t$ in Fig. \ref{trad_mc}, is high. This high sampling frequency leads to partial `cancellation' of the (drag) perturbation effects, resulting in unrealistically small orbit errors. Explanation of the `cancellation' effect - let us consider a hypothetical case of $\Delta t=1s$ and a constant density field with normal distribution $\mathcal{N}(10^{-12} kg/m^3,10^{-13}kg/m^3)$. A series of sampled density values at four consecutive time steps $0s$, $1s$, $2s$, $3s$ can be $.94 \times 10^{-12} kg/m^3$, $1.07 \times 10^{-12} kg/m^3$, $.98 \times 10^{-12} kg/m^3$, and $1.01 \times 10^{-12} kg/m^3$, respectively. The drag force with a density greater than the mean value will increase the along-track error, and the drag force with a density smaller than the mean value will decrease the along-track error. Without a sufficiently large $\Delta t$, the two perturbation effects partially cancel out each other. Such behavior is unrealistic as actual density values have spatiotemporal correlation, i.e., the change in density should be more gradual. In the later part of this paper, we simulate an orbit uncertainty propagation scenario using the traditional Monte Carlo method.

\subsection{Modified Monte Carlo Simulation Techniques for Density Correlation}
The atmospheric density is correlated in both time and space. The sampling procedure for the traditional Monte Carlo scheme needs to be modified to correctly capture the effect of this correlation on the evolution of orbital state uncertainty. Here, we present two Monte Carlo schemes based on the sampling of so-called `bias' factor $\kappa$. If $\kappa_i$ is a sampled value of the bias factor at any point in the orbit for a Monte Carlo run, then the corresponding density sample is $\rho_i=\mu_{\rho_i} + \kappa_i \sigma_{\rho_i}$, where the mean density $\mu_{\rho_i}$ and the standard deviation $\sigma_{\rho_i}$ are obtained from either HASDM-ML-DP or CHAMP-ML-DP or MSIS-UQ-DP density model. When computed across all Monte Carlo runs, the bias factor needs to have the following properties for each epoch of the orbit propagation - (a) $E[\kappa]$ is roughly equal to zero, (b) $Var[\kappa]$ is roughly equal to unity, where $E[\cdot]$ represents the expected value and $Var[\cdot]$ represents the variance. These two desirable properties of the bias factor ensures that the machine learning model predicted moments are preserved, i.e., $E[\rho]=E[\mu_{\rho_i} + \kappa \sigma_{\rho_i}]$ is roughly equal to the model predicted mean $\mu_{\rho_i}$ and $Var[\rho]=Var[\mu_{\rho_i} + \kappa \sigma_{\rho_i}]$ is roughly equal to the model predicted variance $\sigma_{\rho_i}^2$.

\begin{figure*}[hbt!]
\centering
\includegraphics[width=.9\textwidth]{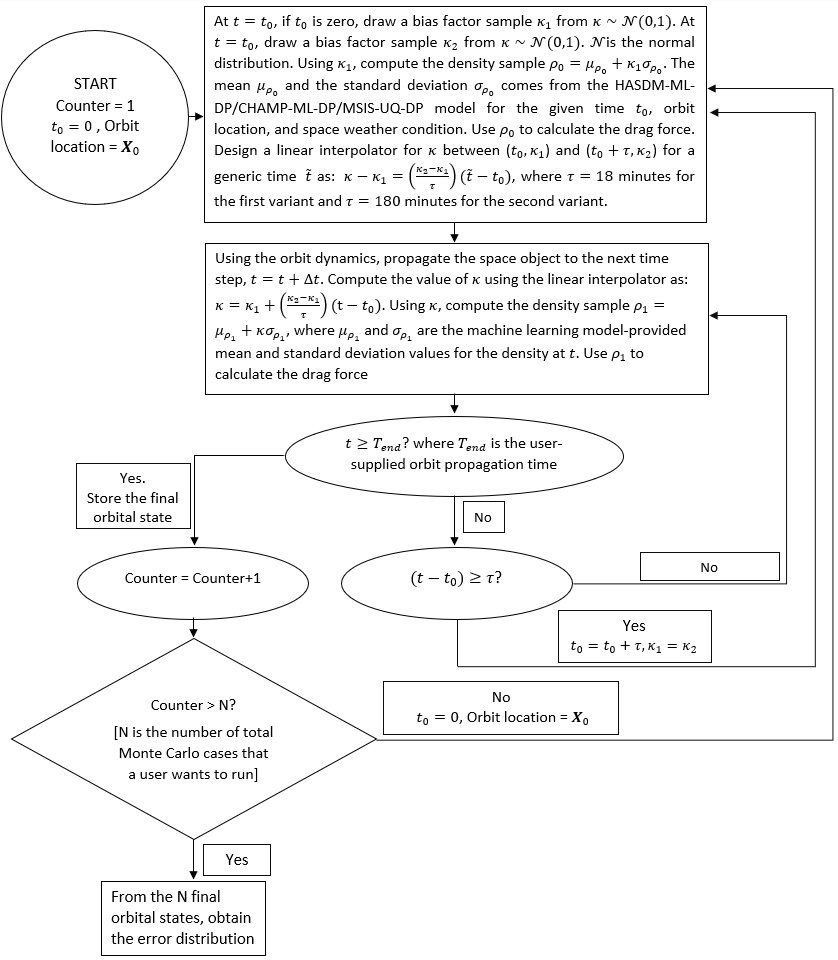}
\caption{Modified Monte Carlo method 1 for orbit uncertainty propagation.}
\label{mod_mc_1}
\end{figure*}

In the first proposed Monte Carlo method, $\kappa$ is sampled from a standard normal distribution. Two variants of this method are implemented - (1) first variant: for each Monte Carlo run, we sample $\kappa$ every 18 minutes and interpolate the value of $\kappa$ in between, (2) second variant: for each Monte Carlo run, we sample $\kappa$ every 180 minutes and interpolate the value in between. Detailed schematic of the first proposed Monte Carlo method is shown in Fig. \ref{mod_mc_1}.

\begin{figure*}[hbt!]
\centering
\includegraphics[width=.9\textwidth]{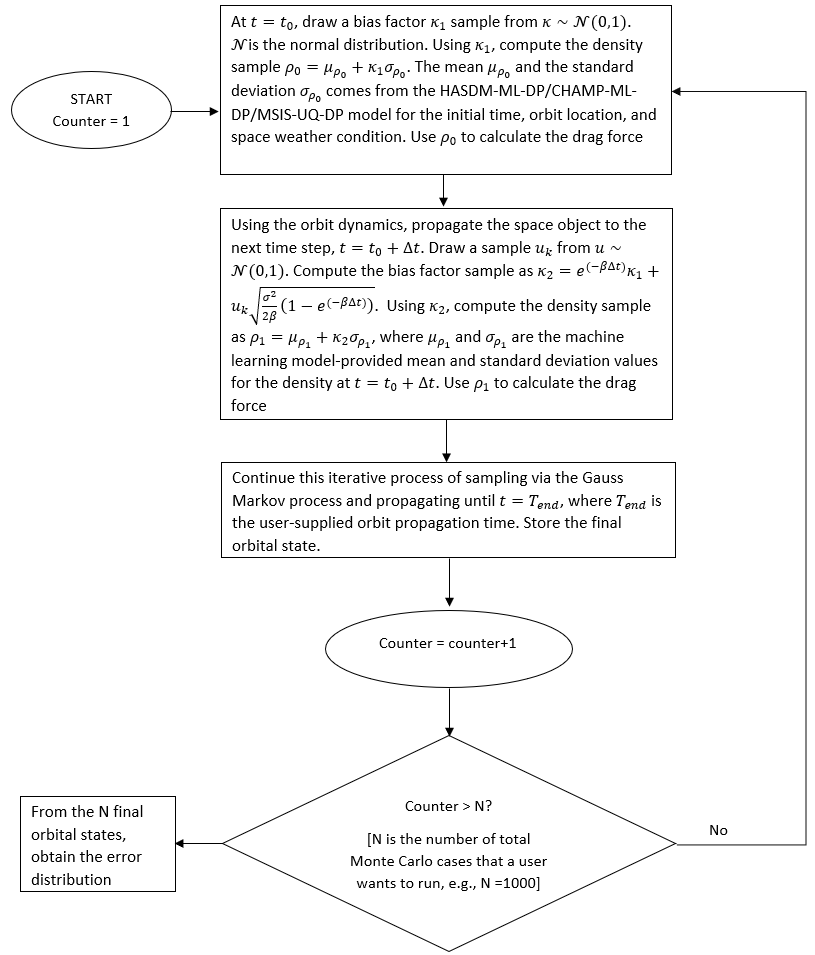}
\caption{Modified Monte Carlo method 2 for orbit uncertainty propagation.}
\label{mod_mc_2}
\end{figure*}

In the second proposed Monte Carlo method, $\kappa$ is sampled from a first order Gauss-Markov process \citep{tapley}:

\begin{subequations}
\begin{equation}
    \kappa(t+\Delta t) = \exp{(-\beta \Delta t)} \kappa(t) + u_k(t) \sqrt{\frac{\sigma^2}{2\beta}\big(1-\exp{(-2\beta \Delta t})\big)}
\end{equation}
\begin{equation}
    \beta = - \frac{\ln{0.5}}{\tau}
\end{equation}
\end{subequations}
where $u_k(t)$ is a random number sampled from the standard normal distribution. The factor $(\sigma^2/(2\beta))$, which represents the steady-state variance of $\kappa$, is taken to be unity. The parameter $\tau$ is the ``half-life'' and governs the rate at which the auto-correlation fades. We implement two variants of the second Monte Carlo method - (1) half-life $\tau$ is taken to be 18 minutes, (2) half-life $\tau$ is taken to be 180 minutes. The value of 18 minutes or 180 minutes is taken from literature \citep{McLaughlin2012}. Figure \ref{mod_mc_2} shows the detailed schematic for the second Monte Carlo approach.

\begin{figure*}[hbt!]
\centering
\includegraphics[width=.9\textwidth]{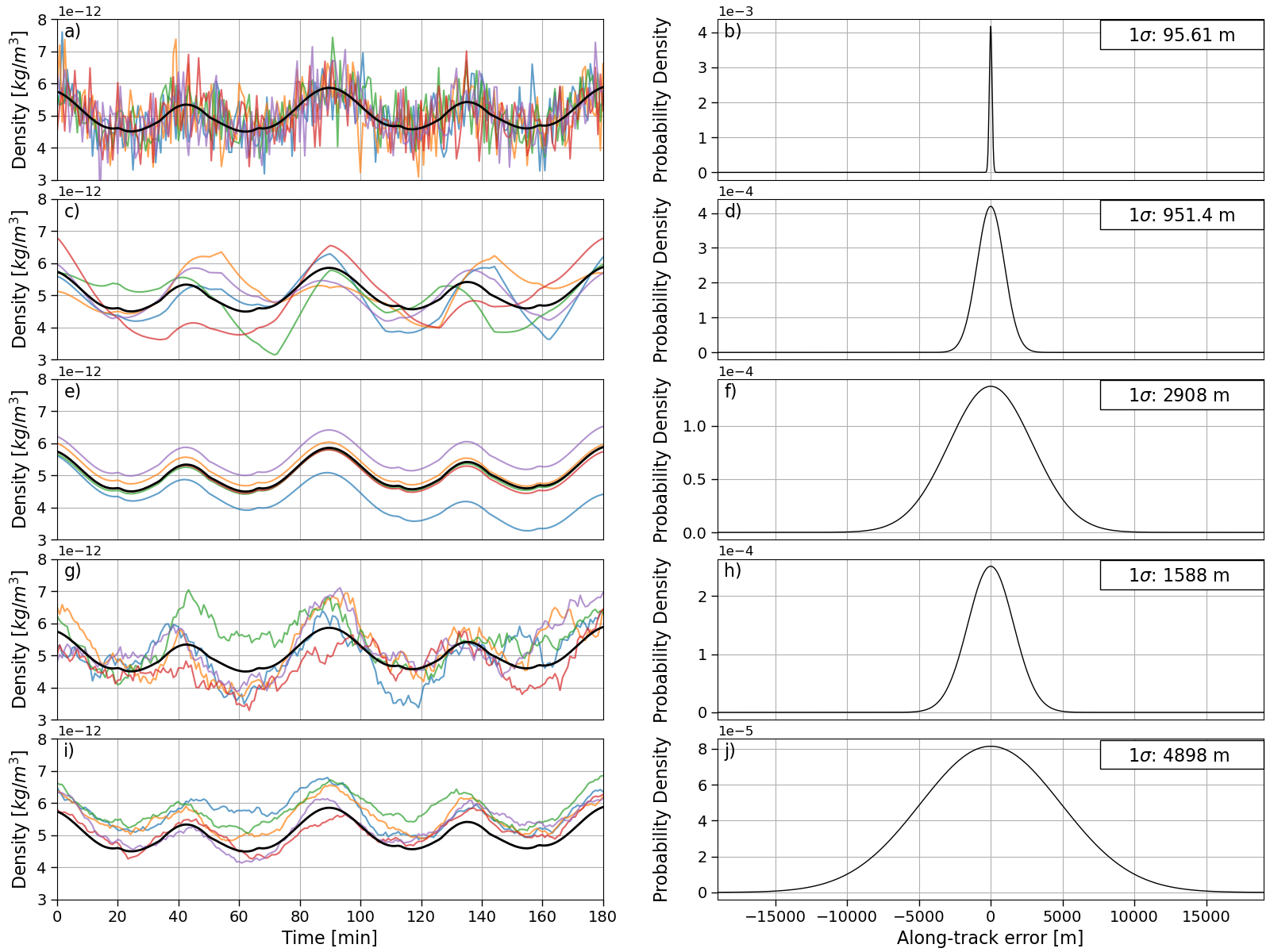}
\caption{Comparison of traditional and modified Monte Carlo techniques. Traditional Monte Carlo approach is shown in (a), (b), modified Monte Carlo method 1 with $\kappa$ = 18 minutes is shown in (c), (d), modified Monte Carlo method 1 with $\kappa$ = 180 minutes is shown in (e), (f), modified Monte Carlo method 2 with half-life = 18 minutes is shown in (g), (h), modified Monte Carlo method 2 with half-life = 180 minutes is shown in (i), (j).}
\label{compare_MCmethods}
\end{figure*}

We simulate a 3-day orbit propagation using various Monte Carlo schemes discussed so far to examine the along-track position error between the mean orbit (i.e., the orbit propagated with mean density) and the Monte Carlo runs. For each Monte Carlo method, we use a total of 1000 Monte Carlo iterations. The initial epoch for the simulation is taken as 01:00:00 UTC, September 07, 2002, which corresponds to a geomagnetic storm from the solar cycle 23. We use a high-inclination LEO orbit and assume no initial uncertainty in the position or velocity. The initial orbit state is taken as $\boldsymbol X_0$ = [3782900.7032 m, -5441600.6779 m, -1420075.1327 m, -606.6600 m/s,  1539.2559 m/s, -7488.3946 m/s] in the Earth-centered inertial (ECI) frame. Apart from the central Earth gravity, only the dominant $J_2$ and atmospheric drag perturbations are considered for the orbit propagation, where the atmospheric density is modeled using the stochastic HASDM-ML-DP model. The object is assumed to be spherically symmetric with a cross-sectional area-to-mass ratio (AMR) value of .0015 $m^2/kg$ and drag coefficient $C_{D_{HASDM}}$ value of 3.0912. Figure \ref{compare_MCmethods} shows a comparison of the different Monte Carlo methods. Figures \ref{compare_MCmethods}(a), \ref{compare_MCmethods}(c), \ref{compare_MCmethods}(e), \ref{compare_MCmethods}(g), and \ref{compare_MCmethods}(i) show the density values for the first three hours of the propagation. The bold black curve in the density plots is the mean density curve, and the five colored curves correspond to the first five Monte Carlo iterations. Figures \ref{compare_MCmethods}(b), \ref{compare_MCmethods}(d), \ref{compare_MCmethods}(f), \ref{compare_MCmethods}(h), and \ref{compare_MCmethods}(j) show the normal PDF for the along-track error at the end of three-day propagation. Figures \ref{compare_MCmethods}(a), \ref{compare_MCmethods}(b) correspond to the traditional Monte Carlo method, Figs. \ref{compare_MCmethods}(c), \ref{compare_MCmethods}(d) correspond to the modified Monte Carlo method 1 with $\kappa$ = 18 minutes, Figs. \ref{compare_MCmethods}(e), \ref{compare_MCmethods}(f) correspond to the modified Monte Carlo method 1 with $\kappa$ = 180 minutes, Figs. \ref{compare_MCmethods}(g), \ref{compare_MCmethods}(h) correspond to the modified Monte Carlo method 2 with a half-life of 18 minutes, and Figs. \ref{compare_MCmethods}(i), \ref{compare_MCmethods}(j) correspond to the modified Monte Carlo method 2 with a half-life of 180 minutes. For the traditional Monte Carlo method, as seen in Fig. \ref{compare_MCmethods}(a), the density variations have little to no spatiotemporal correlation, resulting in a small standard deviation of 95.61 m for the along-track error. As $\kappa$ increases for the modified Monte Carlo method 1, the along-track error (standard deviation value) increases from 951.4 m to 2908 m, as a larger $\kappa$ leads to a stronger spatiotemporal correlation of the density, allowing the orbital errors to grow more in between two sampling times. Similarly, for the modified Monte Carlo method 2, as the half-life increases from 18 minutes to 180 minutes, the along-track error (standard deviation value) increases from 1588 m to 4898 m. A larger half-life means a stronger spatiotemporal density correlation. There is no way to confirm the correct approach, but based on the probailistic density variations (Fig. \ref{compare_MCmethods} (g)) and resulting position distribution (Fig. \ref{compare_MCmethods} (h)), we find Monte Carlo method 2 with an 18-minute half-life to be the most realistic. This approach is used for the remainder of this work.

\section{The Consider Covariance Sigma Point (CCSP) Filter}
The Monte Carlo simulation method is probably the most well-known method for orbit uncertainty propagation. With a sufficiently large number of samples, the method can capture the evolution of any higher-order moments. However, the computing costs are substantial, especially if the goal is to propagate uncertainty in tens of thousands of catalog objects for conjunction assessment over a period of several days or more. An alternative method for orbit uncertainty propagation - one that is computationally cheap and often used in space operations - is the extended Kalman filter (EKF) \citep{10.1115/1.3662552, Bishop_EKF} for linear uncertainty propagation. For highly non-linear systems, the unscented Kalman filter (UKF) \citep{10.1117/12.280797, 1271397, 882463} is a more accurate and convenient alternative to the prevalent EKF because it does not make any linearization approximations, nor does it require the computation of the Jacobian matrices. The UKF relies on non-linear propagation of a select number of the so-called sigma points \citep{10.1117/12.280797}, which are carefully selected to capture the first two moments. The Kalman filtering techniques typically have two steps - a propagation step and an update step. The computations in the update step rely on sensor measurements. Our filtering discussions are limited to the propagation step since the investigations in this paper concern with future predictions in the absence of any measurements.  

There is no direct method to translate the epistemic uncertainty predicted by the HASDM-ML-DP/CHAMP-ML-DP/MSIS-UQ-DP models in atmospheric density to the uncertainty in orbital states (position, velocity) under a traditional EKF/UKF framework. A consider covariance analysis \citep{tapley} based propagation of sigma points can address the issue of translating atmospheric density model uncertainty to uncertainty in state estimates. In this paper, we will refer to such a framework as the consider covariance sigma point (CCSP) filter. The CCSP filter originates from the work by \cite{lisano2006nonlinear}. We are now going to detail the methodological procedure for the implementation of the CCSP filter for orbit uncertainty propagation for the first two time steps.

The state $\boldsymbol{X}$ of interest for our orbit dynamical equations consists of the space object position and velocity, augmented by the ``consider parameter'':

\begin{equation}
    \boldsymbol{X} =  
        \begin{bmatrix}
            x & y & z & v_x & v_y & v_z & \rho 
        \end{bmatrix}^T
\end{equation}
where $x$, $y$, $z$ are the Cartesian position coordinates, $v_x$, $v_y$, $v_z$ are the Cartesian velocity coordinates, and $\rho$ is the atmospheric density.

\subsection{Computations for \texorpdfstring{$t_0$}{}}
At the initial time (i.e., $t=0$ or $t_0$), we assume (almost) no uncertainty in the position and velocity (modeled by numbers $\epsilon_{i}$'s that are very small and arbitrarily close to zero), and the density uncertainty $\sigma_{\rho_0}^2$ is estimated from the HASDM-ML-DP/CHAMP-ML-DP/MSIS-UQ-DP models. We assume the initial cross-correlation between different states to be zero. Mathematically, we have the following equations for the initial mean state and the covariance matrix:
\begin{subequations}
\begin{equation}
        \boldsymbol{X}_{t=0} =  
        \begin{bmatrix}
            x_0 & y_0 & z_0 & v_{x_0} & v_{y_0} & v_{z_0} & \rho_0 
        \end{bmatrix}^T
\end{equation}
\begin{equation}
    P_{\boldsymbol{X}_{t=0}} = Diag\bigg(
            \begin{bmatrix}
            \epsilon_x^2 & \epsilon_y^2 & \epsilon_z^2 & \epsilon_{v_x}^2 & \epsilon_{v_y}^2 & \epsilon_{v_z}^2 & \sigma_{\rho_0}^2
        \end{bmatrix}
    \bigg)
\end{equation}
\end{subequations}
where \textit{Diag} represents the diagonal matrix. As a useful rule of thumb, $\epsilon_{v_i}^2$'s are smaller than $\epsilon_{x}^2$, $\epsilon_{y}^2$, $\epsilon_{z}^2$.

To facilitate the computation of the sigma points, we compute a scaled lower-triangular block-Cholesky decomposition \citep{golub, lisano2006nonlinear} of the initial covariance matrix as:

\begin{equation}
    S_{\boldsymbol{X}_{t=0}} = 
    \begin{bmatrix}
    \sqrt{\lambda_1+n} \;\; Chol([P_{\boldsymbol{X}_{t=0}}]_{(1:n),(1:n)}) & \textbf{0}_{n \times p} \\
    \textbf{0}_{p \times n} & \sqrt{\lambda_2+p} \;\; \sigma_{\rho_0}
    \end{bmatrix}
\end{equation}
where \textit{Chol$(\cdot)$} computes the Cholesky decomposition of the argument matrix and can be computed using Python's Numpy package (NumPy linear algebra function \textit{numpy.linalg.cholesky}). The notation $[\cdot]_{(1:n),(1:n)}$ denotes the submatrix consisting of the first $n$ rows and first $n$ columns of the argument matrix. Parameter $n$ is the dimension of the non-augmented state, i.e., $n = 6$, parameter $p$ is the dimension of the consider parameter, i.e., $p = 1$, $\lambda _1$ and $\lambda _2$ are scaling parameters such that $\lambda_1+n=3$ and $\lambda_2+p=3$, $\textbf{0}_{n \times p}$ denotes the $n \times p$ all-zero matrix, and $\textbf{0}_{p \times n}$ denotes the $p \times n$ all-zero matrix.

Using the $S_{\boldsymbol{X}_{t=0}}$ matrix, two sets of sigma points are computed. The first set, comprising $2n+1$ sigma points (13 sigma points), is given by:

\begin{subequations}
\begin{equation}
    \mathcal{X}_{(i)_{t=0}} = \boldsymbol{X}_{t=0}\;\;\;\;\;\;\;\;\;\;\;\;\;\;\;\text{where}\;\;i = 0
\end{equation}
\begin{equation}
    \mathcal{X}_{(i)_{t=0}} = \boldsymbol{X}_{t=0} + \big[S_{\boldsymbol{X}_{t=0}}\big]_i\;\;\;\;\;\;\text{where}\;\;  i = 1:n
\end{equation}
\begin{equation}
    \mathcal{X}_{(i+n)_{t=0}} = \boldsymbol{X}_{t=0} - \big[S_{\boldsymbol{X}_{t=0}}\big]_i\;\;\;\;\text{where}\;\; i = 1:n
\end{equation}
\end{subequations}
where $[\cdot]_i$ represents the $i^{th}$ column of the argument matrix. 

The second set, comprising $2p+1$ sigma points (3 sigma points), accounts for the uncertainty in the consider parameter $\rho$ and is given by:

\begin{subequations}
\begin{equation}
    \mathcal{x}_{(i)_{t=0}} = \boldsymbol{X}_{t=0}\;\;\;\;\;\;\;\;\;\;\;\;\;\;\;\;\;\;\text{where}\;\;i = 0
\end{equation}
\begin{equation}
    \mathcal{x}_{(i)_{t=0}} = \boldsymbol{X}_{t=0} + \big[S_{\boldsymbol{X}_{t=0}}\big]_{n+i}\;\;\;\;\;\;\text{where}\;\;  i = 1:p
\end{equation}
\begin{equation}
    \mathcal{x}_{(i+p)_{t=0}} = \boldsymbol{X}_{t=0} - \big[S_{\boldsymbol{X}_{t=0}}\big]_{n+i}\;\;\;\;\text{where}\;\; i = 1:p
\end{equation}
\end{subequations}
where, for our orbit propagation problem, $\big[S_{\boldsymbol{X}_{t=0}}\big]_{n+i}=\big[S_{\boldsymbol{X}_{t=0}}\big]_{n+p}$ is the last column of $S_{\boldsymbol{X}_{t=0}}$ matrix.

The first set of 13 sigma points and the second set of three sigma points are then propagated from the initial time to the next time step ($t = t_0+\Delta t$) using perturbed two-body dynamics to obtain $\mathcal{X}_{(i)_{t=t_0+\Delta t}}$ and $\mathcal{x}_{(i)_{t=t_0+\Delta t}}$, respectively. Only the dominant $J_2$ and atmospheric drag perturbations are considered in this work.

\subsection{Computations for \texorpdfstring{$t_0+\Delta t$}{}}
At $t=t_0+\Delta t$, the first step involves estimating the mean and the covariance matrix from the propagated sigma points using a weighted averaging method. The mean and the covariance matrix from the first set of sigma points are computed as:

\begin{subequations}
\begin{equation}
            \boldsymbol{X}_{\mathcal{X}, t=t_0+\Delta t} =  
            \sum_{i=0}^{2n}\bigg[w_i^{(mean)}\mathcal{X}_{(i)_{t=t_0+\Delta t}}\bigg]
\end{equation}
\begin{equation}
    P_{\boldsymbol{X}_{\mathcal{X}, t=t_0+\Delta t}} = 
    \sum_{i=0}^{2n}\bigg[w_i^{(cov)}\big(\mathcal{X}_{(i)_{t=t_0+\Delta t}}-\boldsymbol{X}_{\mathcal{X}, t=t_0+\Delta t}\big)\big(\mathcal{X}_{(i)_{t=t_0+\Delta t}}-\boldsymbol{X}_{\mathcal{X}, t=t_0+\Delta t}\big)^T\bigg]
\end{equation}
\end{subequations}
where the weights are derived from the standard UKF framework and are obtained as:

\begin{subequations}
\begin{equation}
    w_0^{(mean)} = w_0^{(cov)} = \frac{\lambda_1}{n + \lambda_1}
\end{equation}
\begin{equation}
    w_i^{(mean)} = w_i^{(cov)} = \frac{1}{2(n + \lambda_1)} \;\;\;\;\;\;\text{where}\;\;  i = 1:2n
\end{equation}
\end{subequations}

Similarly, we compute the mean and the covariance matrix for the second set of $2p+1$ propagated sigma points using:

\begin{subequations}
\begin{equation}
            \boldsymbol{X}_{\mathcal{x}, t=t_0+\Delta t} =  
            \sum_{i=0}^{2p}\bigg[\omega_i^{(mean)}\mathcal{x}_{(i)_{t=t_0+\Delta t}}\bigg]
\end{equation}
\begin{equation}
    P_{\boldsymbol{X}_{\mathcal{x}, t=t_0+\Delta t}} = 
    \sum_{i=0}^{2p}\bigg[\omega_i^{(cov)}\big(\mathcal{x}_{(i)_{t=t_0+\Delta t}}-\boldsymbol{X}_{\mathcal{x}, t=t_0+\Delta t}\big)\big(\mathcal{x}_{(i)_{t=t_0+\Delta t}}-\boldsymbol{X}_{\mathcal{x}, t=t_0+\Delta t}\big)^T\bigg]
\end{equation}
\end{subequations}
with weights obtained as:

\begin{subequations}
\begin{equation}
    \omega_0^{(mean)} = \omega_0^{(cov)} = \frac{\lambda_2}{p + \lambda_2}
\end{equation}
\begin{equation}
    \omega_i^{(mean)} = \omega_i^{(cov)} = \frac{1}{2(p + \lambda_2)} \;\;\;\;\;\;\text{where}\;\;  i = 1:2p
\end{equation}
\end{subequations}
Following the computation of the consider covariance matrix $P_{\boldsymbol{X}_{\mathcal{x}, t=t_0+\Delta t}}$, we compute the $n\times p$ cross-correlation matrix as:

\begin{equation}
    P_{\boldsymbol{cross}_{t=t_0+\Delta t}} = 
    \begin{bmatrix}
            P_{\boldsymbol{X}_{\mathcal{x}, t=t_0+\Delta t}}
    \end{bmatrix}_{(1:n), (n+1:n+p)}
\end{equation}
where the notation $[\cdot]_{(1:n), (n+1:n+p)}$ denotes the submatrix consisting of the first $n$ rows and last $p$ columns of the argument matrix. 

From the cross-correlation matrix $P_{\boldsymbol{cross}_{t=t_0+\Delta t}}$, one can compute the additive uncertainty resulting from the inclusion of the consider parameter $\rho$ as:

\begin{equation}
    dP_{t=t_0+\Delta t} = P_{\boldsymbol{cross}_{t=t_0+\Delta t}}\Bigg(\frac{1}{\sigma_{\rho_1}^2}\Bigg) P_{\boldsymbol{cross}_{t=t_0+\Delta t}}^T
\end{equation}
where $\sigma_{\rho_1}^2$ is the atmospheric density uncertainty estimate from the HASDM-ML-DP/CHAMP-ML-DP models computed at $t=t_0+\Delta t$ at the mean position indicated by the first three coordinates of the vector $\boldsymbol{X}_{\mathcal{X}, t=t_0+\Delta t}$.

The covariance matrix updated for the uncertainty in the consider parameter is then given as:

\begin{equation}
    P_{\boldsymbol{X}_{t=t_0+\Delta t}} = 
    \begin{bmatrix}
            [P_{\boldsymbol{X}_{\mathcal{X}, t=t_0+\Delta t}}]_{(1:n),(1:n)} +   [dP_{t=t_0+\Delta t}]_{(1:n),(1:n)} & P_{\boldsymbol{cross}_{t=t_0+\Delta t}} \\
            P_{\boldsymbol{cross}_{t=t_0+\Delta t}}^T & \sigma_{\rho_1}^2
    \end{bmatrix}
\end{equation}

A scaled lower-triangular block-Cholesky decomposition of the matrix $P_{\boldsymbol{X}_{t=t_0+\Delta t}}$ is then obtained as:

\begin{equation}
        S_{\boldsymbol{X}_{t=t_0+\Delta t}} = 
    \begin{bmatrix}
     Chol(A) & \textbf{0}_{n \times p} \\
    \bigg(\frac{1}{\sqrt{\lambda_2+p} \;\; \sigma_{\rho_1}}\bigg) P_{\boldsymbol{cross}_{t=t_0+\Delta t}}^T  & \sqrt{\lambda_2+p} \;\; \sigma_{\rho_1}
    \end{bmatrix}
\end{equation}
where the $n \times n$ matrix $A$ is given by:

\begin{subequations}
\begin{equation}
    A = (\lambda_1 + n) [P_{\boldsymbol{X}_{t=t_0+\Delta t}}]_{(1:n), (1:n)} - B^T B
\end{equation}
\begin{equation}
    B = \bigg(\frac{1}{\sqrt{\lambda_2+p} \;\; \sigma_{\rho_1}}\bigg) P_{\boldsymbol{cross}_{t=t_0+\Delta t}}^T
\end{equation}
\end{subequations}

Thereafter, using the $S_{\boldsymbol{X}_{t=t_0+\Delta t}}$ matrix, two sets of sigma points are computed. The first set of sigma points is given by:

\begin{subequations}
\begin{equation}
    \mathcal{X}_{(i)_{t=t_0+\Delta t}} = \boldsymbol{X}_{\mathcal{X}, t=t_0+\Delta t}\;\;\;\;\;\;\;\;\;\;\;\;\;\;\;\;\;\;\text{where}\;\;i = 0
\end{equation}
\begin{equation}
    \mathcal{X}_{(i)_{t=t_0+\Delta t}} = \boldsymbol{X}_{\mathcal{X}, t=t_0+\Delta t} + \big[S_{\boldsymbol{X}_{t=t_0+\Delta t}}\big]_i\;\;\;\;\;\;\text{where}\;\;  i = 1:n
\end{equation}
\begin{equation}
    \mathcal{X}_{(i+n)_{t=t_0+\Delta t}} = \boldsymbol{X}_{\mathcal{X}, t=t_0+\Delta t} - \big[S_{\boldsymbol{X}_{t=t_0+\Delta t}}\big]_i\;\;\;\text{where}\;\;  i = 1:n
\end{equation}
\end{subequations}
The second set of sigma points is given by:

\begin{subequations}
\begin{equation}
    \mathcal{x}_{(i)_{t=t_0+\Delta t}} = \boldsymbol{X}_{\mathcal{X}, t=t_0+\Delta t}\;\;\;\;\;\;\;\;\;\;\;\;\;\text{where}\;\;i = 0
\end{equation}
\begin{equation}
    \mathcal{x}_{(i)_{t=t_0+\Delta t}} = \boldsymbol{X}_{\mathcal{X}, t=t_0+\Delta t} + \big[S_{\boldsymbol{X}_{t=t_0+\Delta t}}\big]_{n+i}\;\;\;\;\;\;\text{where}\;\;  i = 1:p
\end{equation}
\begin{equation}
    \mathcal{x}_{(i+p)_{t=t_0+\Delta t}} = \boldsymbol{X}_{\mathcal{X}, t=t_0+\Delta t} - \big[S_{\boldsymbol{X}_{t=t_0+\Delta t}}\big]_{n+i}\;\;\;\text{where}\;\; i = 1:p
\end{equation}
\end{subequations}
Both sets of sigma points are then propagated to the next time step.

\section{Ensemble Modeling for Orbit Uncertainty Quantification}
There is no evidence that any single atmospheric density model is always more accurate than alternative density models under all space weather conditions \citep{VALLADO2014141}. As a result, drawing firm conclusions about orbital state uncertainties caused by atmospheric density uncertainties from a single density model is not recommended. The current study, therefore, proposes a multi-model ensemble approach where the final orbit state PDF is expressed as a combination of orbit state PDFs resulting from the individual HASDM-ML-DP, CHAMP-ML-DP, and MSIS-UQ-DP thermospheric density models. Although we use just three stochastic density models in this study, the multi-model ensemble approach for orbit uncertainty quantification can be expanded to include any number of atmospheric density models. For completeness, readers should be aware of the alternative input-based ensemble modeling technique in which a single model is used, but the starting condition or driver input is perturbed to produce many outputs \citep{murray2018}, which are then aggregated to provide the final uncertainty distribution. Next, we turn attention to the implementation of the proposed ensemble approach, which we describe through a dummy example.

Let us consider an LEO satellite \textbf{A} whose initial orbital state (Cartesian position and velocity components) is given to us. The given state information consists of the initial mean and the covariance matrix. Furthermore, let us assume that the satellite has a constant cross-sectional area and a constant mass and that the atmosphere co-rotates with the Earth. As a first step for the ensemble approach, using the HASDM-ML-DP density model and the CCSP filter (or the Monte Carlo approach), let \textbf{A} be propagated over a user-defined period of choice. A constant value of drag coefficient $C_{D_{HASDM}}$ derived from a physical model is used for the propagation. In operational setups, ballistic coefficient (or drag coefficient if the cross-sectional area and mass are known) and atmospheric density are often estimated simultaneously to match the satellite observations (refer Eq.~\eqref{drag_accn}), if available. In other words, if $\rho_1$ and $\rho_2$ are the density estimates from two different atmospheric density models and if $CD_1$ is the drag coefficient estimate corresponding to the first density model, then the drag coefficient estimate $CD_2$ corresponding to the second density model is such that $\rho_1 CD_1=\rho_2 CD_2$. Even in absence of any satellite measurements, it is a desirable practice to occasionally ``debias'' $CD_2$ using $CD_2=(\rho_1/\rho_2) CD_1$. This little detour to explain the concept of debiasing is essential for the next step. As a second step for the ensemble approach, let us propagate \textbf{A} using the CHAMP-ML-DP density model for the same initial condition and the same period of time as earlier. The corresponding drag coefficient $C_{D_{CHAMP}}$ is obtained from a debiasing scheme based on the average ratio of densities predicted by the HASDM-ML-DP and CHAMP-ML-DP models along the orbit for a user-defined period before the initial epoch. As a third step for the ensemble approach, we propagate \textbf{A} using the MSIS-UQ-DP density model for the same initial condition and time period as earlier. The corresponding drag coefficient $C_{D_{MSIS}}$ is obtained by a debiasing scheme based on the average ratio of densities predicted by the HASDM-ML-DP and MSIS-UQ-DP models along the orbit for a user-defined period before the initial epoch.

Let $\boldsymbol\mu_1$ and $\boldsymbol P_1$ be the mean position and the covariance matrix representing positional uncertainty at the end of the propagation period using HASDM-ML-DP model. Let $\boldsymbol\mu_2$, $\boldsymbol P_2$ represent the mean position and the covariance matrix at the end of the propagation period following CHAMP-ML-DP model. And, let $\boldsymbol\mu_3$, $\boldsymbol P_3$ represent the mean position and the covariance at the end of the propagation using MSIS-UQ-DP model. In our ensemble formulation, the final PDF for the orbital position is given by an equally weighted Gaussian mixture as:

\begin{equation}
    p ({X}; \boldsymbol{\mu_1}, \boldsymbol P_1, \boldsymbol\mu_2, \boldsymbol P_2,\boldsymbol\mu_3, \boldsymbol P_3) = \frac{1}{3}\mathcal{N}({X}; \boldsymbol\mu_1, \boldsymbol P_1) + \frac{1}{3}\mathcal{N}({X}; \boldsymbol\mu_2, \boldsymbol P_2) +
    \frac{1}{3}\mathcal{N}({X}; \boldsymbol\mu_3, \boldsymbol P_3)
\end{equation}
where $\mathcal{N}(\boldsymbol\mu_i, \boldsymbol P_i)$ represents the multivariate normal distribution with mean $\boldsymbol\mu_i$ and covariance $\boldsymbol P_i$.

\section{Results}
Space weather has a strong influence on the propagation and prediction of orbital states. In this section, the evolution of orbital state uncertainty is investigated using the ensemble approach under four different space weather conditions:
\begin{enumerate}
    \item Case-I: a geomagnetic storm during solar maximum (circa September 07, 2002), which we refer to as `solar-max-storm'.
    \item Case-II: a non-storm period during solar minimum (circa December 02, 2009), which we refer to as `solar-min'.
    \item Case-III: a simulated solar maximum scenario (circa September 07, 2002) where we keep the geomagnetic model drivers constant at global mean values and let all other drivers (e.g., solar flux values) have the same variations as that of case-I. The global mean values for the geomagnetic model drivers are obtained from the machine learning models' training data. We refer to this case as `solar-max-simulated'.
    \item Case-IV: a simulated non-storm solar minimum case (circa December 02, 2009) where we keep the geomagnetic model drivers constant at global mean values and let all other drivers have the same variations as that of case-II. We refer to this case as `solar-min-simulated'.
\end{enumerate}

    A 3-day orbit uncertainty propagation for a high-inclination LEO object is carried out for the four cases in the presence of $J_2$ and atmospheric drag orbital perturbations. The details of the simulations are given in Table \ref{Table1}.
    
    \begin{table*}[htbp]
	\fontsize{10}{10}\selectfont
    \caption{Simulation set-up for the orbit uncertainty propagation using super-ensemble approach.}
   \label{Table1}
        \centering 
   \begin{tabular}{c | c } 
      \hline 
      Parameter &  Value/Details\\
      \hline \hline
      Initial position (ECI) &   [3782900.7032, -5441600.6779, -1420075.1327] m\\
      \hline
      Initial velocity (ECI) &  [-606.6600,  1539.2559, -7488.3946] m/s \\
      \hline
      Propagation period &  259200 s \\      
      \hline
      Object shape/type & Spherical \& symmetric \\            
      \hline
      Cross-sectional AMR  &  .0015 $m^2/kg$ \\
      \hline
      Drag coefficient $C_{D_{HASDM}}$  &  3.0912 \\       
      \hline
      Propagation period for debiasing  & 8 hours \\
      \hline
      Initial epoch for case-I and case-III & 01:00:00 UTC, September 07, 2002\\
      \hline
      Initial epoch for case-II and case-IV & 00:00:00 UTC, December 02, 2009\\
      \hline
      Orbit propagation method & Modified Monte Carlo method 2 with half-life = 18 minutes\\
      \hline
      Number of Monte Carlo iterations & 1000 for each case for each density model\\
      \hline
   \end{tabular}
\end{table*}

\begin{figure*}[hbt!]
\centering
\includegraphics[width=.8\textwidth]{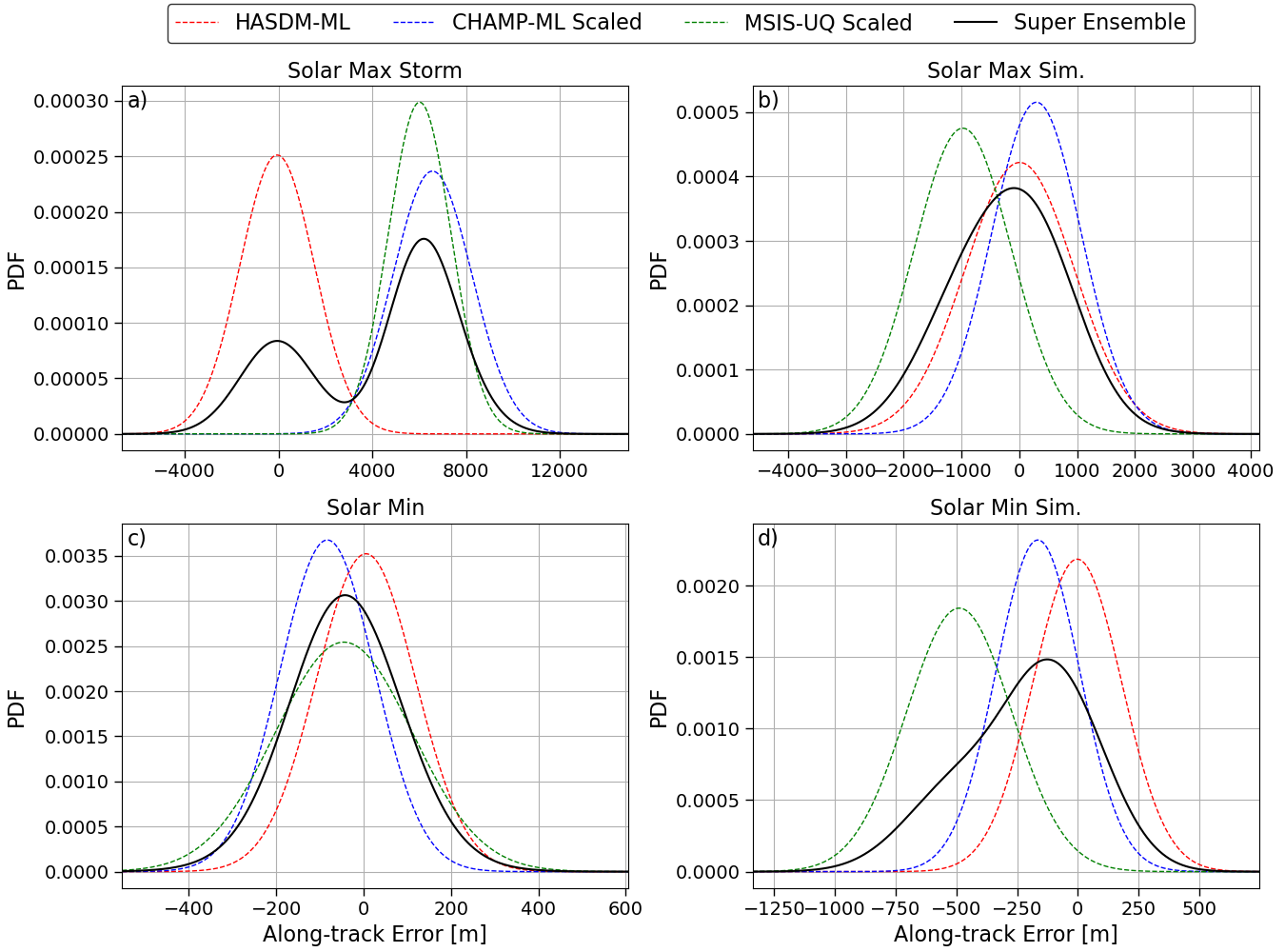}
\caption{Orbit state PDF for the along-track direction using the super ensemble approach. Top-left figure corresponds to case-I (solar-max-storm), bottom-left figure corresponds to case-II (solar-min), top-right figure corresponds to case-III (solar-max-simulated), and the bottom-right figure corresponds to case-IV (solar-min-simulated).}
\label{ensemble_result}
\end{figure*}

In Fig. \ref{ensemble_result}, we show the orbit state PDF for the along-track direction at the end of the 3-day propagation period. Figure \ref{ensemble_result}(a) quantifies the uncertainty in orbit state for the solar-max-storm case, Fig. \ref{ensemble_result}(c) shows the orbit state distribution for the solar-min case, Fig. \ref{ensemble_result}(b) shows the orbit state distribution for the solar-max-simulated case, and Fig. \ref{ensemble_result}(d) shows the PDF for the solar-min-simulated case. In each of the figures, the red dashed curve corresponds to the orbit state PDF for propagation using the HASDM-ML-DP model, the blue dashed curve corresponds to the PDF for propagation using the CHAMP-ML-DP model, and the green dashed curve corresponds to the PDF for propagation using the MSIS-UQ-DP model. The term `scaled' in the title indicates that the drag coefficients for the CHAMP-based and MSIS-based propagations are obtained by scaling the $C_{D_{HASDM}}$ using the debiasing scheme discussed earlier. The bold black curve shows the resultant orbit state PDF from the super ensemble approach. When using more than one stochastic atmospheric density model, an ensemble technique unquestionably yields distributions that are capable of deviating greatly from the normal distribution that would otherwise be obtained. Figure \ref{ensemble_result} shows that the uncertainties for the solar maximum cases are larger compared to the solar minimum cases, highlighting the importance of the space weather condition in uncertainty quantification. For the solar-max-storm case, the ensemble approach leads to almost a bimodal uncertainty distribution, indicating the presence of two distinct regions of high probability for the orbit state. From the plot for the solar-max-simulated case, in the absence of geomagnetic variations, there is a reduced bias between the mean positions predicted by the HASDM-ML-DP and CHAMP-ML-DP models compared to the solar-max-storm case. As the geomagnetic variations are small in the solar minimum period, no such observation is made in comparison between solar-min-simulated and solar-min cases.

\begin{figure*}[hbt!]
\centering
\includegraphics[width=.8\textwidth]{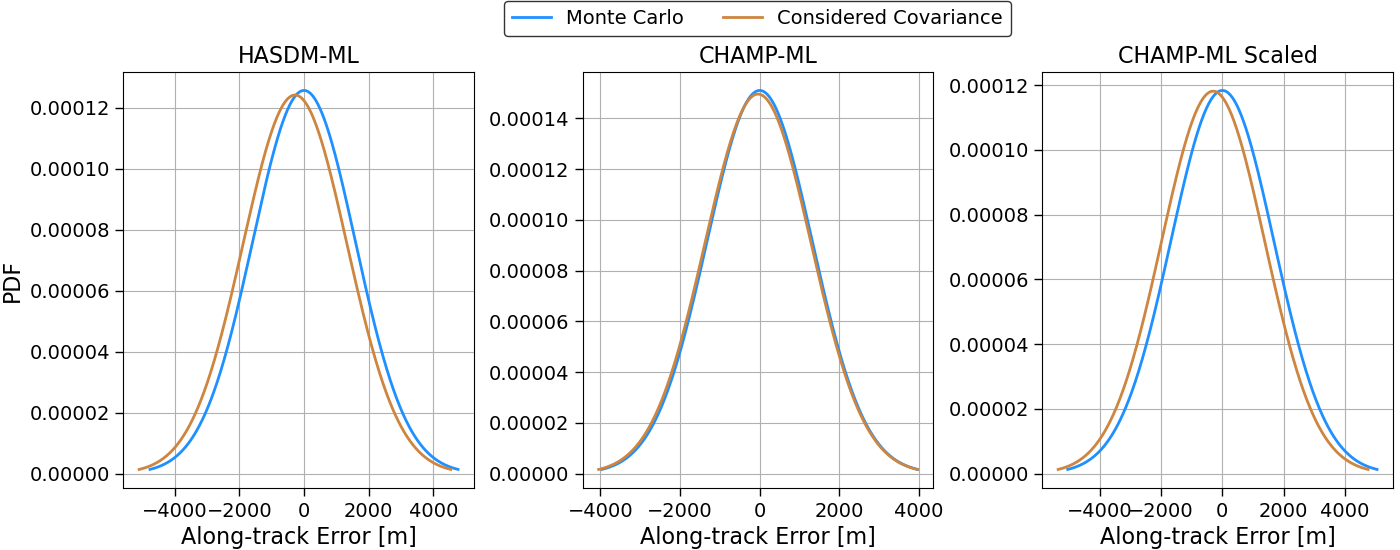}
\caption{Comparison between Monte Carlo and CCSP filter-based methods of orbit uncertainty propagation under solar-max-storm space weather condition.}
\label{compare_MC_CCSP}
\end{figure*}

In addition to the results for the ensemble approach shown in Fig. \ref{ensemble_result}, a comparison between the Monte Carlo approach (modified Monte Carlo method 2 with a half-life of 18 minutes) and consider covariance approach to orbit uncertainty propagation is also demonstrated for a select few density models. Figure \ref{compare_MC_CCSP} shows the orbit state PDF for the along-track direction for case-I, i.e., the solar-max-storm condition. The left figure corresponds to the propagation using the HASDM-ML-DP density model, the middle figure corresponds to the propagation using the CHAMP-ML-DP model using the drag coefficient $C_{D_{HASDM}}$ (i.e., no debiasing is performed), and the right figure corresponds to the propagation using CHAMP-ML-DP model using the drag coefficient obtained from the debiasing scheme. In each of the figures, the blue curve corresponds to the Monte Carlo approach, and the orange curve corresponds to the CCSP filter. Similar to the case of the traditional Monte Carlo approach, if the covariance update rate in the CCSP filter is high ($\Delta t$ is small), the spatiotemporal density correlation is not captured correctly, resulting in unrealistically small uncertainty values. Based upon manual tuning, we use a covariance update rate of 60 minutes for the CCSP filter, where the density field varies according to the selected model between consecutive covariance updates. Clearly, from Fig. \ref{compare_MC_CCSP}, the CCSP filter results and the Monte Carlo results are comparable with the added benefit of the CCSP filter being computationally much faster.

\section{Conclusion}
Correct quantification and propagation of orbital uncertainties are critical for space situational awareness functionalities like the computation of probability of collision and track association. Uncertainty in atmospheric density modeling is one of the primary sources of uncertainty in orbit state prediction. Currently, most researchers propagate orbital uncertainties by considering uncertainty in atmospheric density model drivers and using a single density model.

In this paper, we use machine learning-based models that directly provide the epistemic uncertainty in the atmospheric density prediction. We use three stochastic density models -- HASDM-ML-DP, CHAMP-ML-DP, and MSIS-UQ-DP -- to investigate the effect of atmospheric density uncertainty on the evolution of orbit state probability density function (PDF). The popular and traditional Monte Carlo approach for orbit uncertainty propagation fails to capture the spatiotemporal correlation of the atmospheric density. We, therefore, propose four modified Monte Carlo schemes, the first two based on the sampling and interpolation of the so-called bias factor from a normal distribution and the last two based on the first-order Gauss-Markov process, for orbit uncertainty propagation while successfully capturing the spatiotemporal density correlation. Since Monte Carlo methods are computationally expensive, we also investigate the ``consider covariance sigma point (CCSP)'' filter that can perform orbit uncertainty propagation at a much smaller computational cost.

We propose a new super ensemble approach for predicting orbit state PDF that combines the uncertainty distributions predicted individually by each of the three stochastic density models. The super ensemble approach provides more realistic uncertainty estimates as no single density model is always more accurate across different regions of space and time. The three machine learning-based density models were developed using three different satellite data sources, each having unique advantages and disadvantages. In designing the super ensemble framework, we also ensure that appropriate drag coefficient values are used for each of the three density models. The drag coefficient value for the HASDM-ML-DP model is obtained from a physical model, the drag coefficient value for the CHAMP-ML-DP model is obtained from a debiasing scheme based on the average ratio of densities predicted by the HASDM-ML-DP and CHAMP-ML-DP models, and the drag coefficient value for the MSIS-UQ-DP model is obtained from a debiasing scheme based on the average ratio of densities predicted by the HASDM-ML-DP and MSIS-UQ-DP models.

To test our developed ensemble approach, we simulate a 3-day orbit propagation for a high inclination LEO object under four different space weather condition - (i) a geomagnetic storm during solar maximum, (ii) a non-storm condition with geomagnetic variations set to global mean values during solar maximum, (iii) a non-storm solar minimum period, and, (iv) a non-storm condition with geomagnetic variations set to global mean values during solar minimum. Our study shows that an ensemble approach can result in an orbit state PDF that can deviate significantly from a normal PDF resulting from a single density model. We find that the spread in the PDF or uncertainties is much larger for solar maximum conditions as compared to the solar minimum conditions. Furthermore, we see that the bias or difference between the mean positions predicted by different density models can be significantly influenced by geomagnetic variations that occur during a storm. 

The developed ensemble framework provides a novel and realistic way to model orbit uncertainties for satellite operations and the broader space weather community.

\bibliographystyle{unsrtnat}
\bibliography{main_paper_references}  

\begin{thebibliography}{41}
\providecommand{\natexlab}[1]{#1}
\providecommand{\url}[1]{\texttt{#1}}
\expandafter\ifx\csname urlstyle\endcsname\relax
  \providecommand{\doi}[1]{doi: #1}\else
  \providecommand{\doi}{doi: \begingroup \urlstyle{rm}\Url}\fi

\bibitem[Licata and Mehta(2022)]{Licata2022}
Richard~J. Licata and Piyush~M. Mehta.
\newblock Uncertainty quantification techniques for data-driven space weather
  modeling: thermospheric density application.
\newblock \emph{Scientific Reports}, 12\penalty0 (7256), 2022.
\newblock \doi{10.1038/s41598-022-11049-3}.

\bibitem[Licata et~al.(2022{\natexlab{a}})Licata, Mehta, Weimer, Tobiska, and
  Yoshii]{msisuq_arxiv}
Richard~J. Licata, Piyush~M. Mehta, Daniel~R. Weimer, W.~Kent Tobiska, and Jean
  Yoshii.
\newblock Calibrated and enhanced nrlmsis 2.0 model with uncertainty
  quantification, 2022{\natexlab{a}}.
\newblock URL \url{https://arxiv.org/abs/2208.11619}.

\bibitem[McDowell(2020)]{McDowell_2020}
Jonathan~C. McDowell.
\newblock The low earth orbit satellite population and impacts of the {SpaceX}
  starlink constellation.
\newblock \emph{The Astrophysical Journal}, 892\penalty0 (2):\penalty0 L36,
  2020.
\newblock \doi{10.3847/2041-8213/ab8016}.

\bibitem[Henri(2020)]{Henri2020}
Yvon Henri.
\newblock \emph{The OneWeb Satellite System}, pages 1--10.
\newblock Springer International Publishing, Cham, 2020.
\newblock ISBN 978-3-030-20707-6.
\newblock \doi{10.1007/978-3-030-20707-6_67-1}.

\bibitem[Boley and Byers(2021)]{Boley2021}
Aaron~C. Boley and Michael Byers.
\newblock Satellite mega-constellations create risks in low {Earth} orbit, the
  atmosphere and on {Earth}.
\newblock \emph{Scientific Reports}, 11\penalty0 (10642), 2021.
\newblock \doi{10.1038/s41598-021-89909-7}.

\bibitem[Finley et~al.(2013)Finley, Rose, Nave, Wells, Redfern, Rose, and
  Ruf]{finley2013techniques}
Tiffany Finley, Debi Rose, Kyle Nave, William Wells, Jillian Redfern, Randy
  Rose, and Chris Ruf.
\newblock Techniques for {LEO} constellation deployment and phasing utilizing
  differential aerodynamic drag.
\newblock \emph{2013 AAS/AIAA Astrodynamics Specialist Conference AAS 13-797},
  2013.

\bibitem[Vallado and Finkleman(2014)]{VALLADO2014141}
David~A. Vallado and David Finkleman.
\newblock A critical assessment of satellite drag and atmospheric density
  modeling.
\newblock \emph{Acta Astronautica}, 95:\penalty0 141--165, 2014.
\newblock \doi{10.1016/j.actaastro.2013.10.005}.

\bibitem[Bruinsma et~al.(2021)Bruinsma, Boniface, Sutton, and
  Fedrizzi]{bruinsma2021}
S.~Bruinsma, C.~Boniface, E.K. Sutton, and M.~Fedrizzi.
\newblock Thermosphere modeling capabilities assessment: geomagnetic storms.
\newblock \emph{Journal of Space Weather and Space Climate}, 11\penalty0 (12),
  2021.
\newblock \doi{10.1051/swsc/2021002}.

\bibitem[Hapgood et~al.(2022)Hapgood, Liu, and Lugaz]{2022SW003074}
Mike Hapgood, Huixin Liu, and Noé Lugaz.
\newblock {SpaceX}—sailing close to the space weather?
\newblock \emph{Space Weather}, 20\penalty0 (3), 2022.
\newblock \doi{10.1029/2022SW003074}.

\bibitem[Emmert(2015)]{EMMERT2015773}
J.T. Emmert.
\newblock Thermospheric mass density: A review.
\newblock \emph{Advances in Space Research}, 56\penalty0 (5):\penalty0
  773--824, 2015.
\newblock \doi{10.1016/j.asr.2015.05.038}.

\bibitem[He et~al.(2018)He, Yang, Carter, Kerr, Wu, Deleflie, Cai, Zhang,
  Sagnières, and Norman]{HE201831}
Changyong He, Yang Yang, Brett Carter, Emma Kerr, Suqin Wu, Florent Deleflie,
  Han Cai, Kefei Zhang, Luc Sagnières, and Robert Norman.
\newblock Review and comparison of empirical thermospheric mass density models.
\newblock \emph{Progress in Aerospace Sciences}, 103:\penalty0 31--51, 2018.
\newblock \doi{10.1016/j.paerosci.2018.10.003}.

\bibitem[Storz et~al.(2005)Storz, Bowman, Branson, Casali, and
  Tobiska]{STORZ20052497}
Mark~F. Storz, Bruce~R. Bowman, Major James~I. Branson, Stephen~J. Casali, and
  W.~Kent Tobiska.
\newblock High accuracy satellite drag model {(HASDM)}.
\newblock \emph{Advances in Space Research}, 36\penalty0 (12):\penalty0
  2497--2505, 2005.
\newblock \doi{10.1016/j.asr.2004.02.020}.

\bibitem[Gao et~al.(2020)Gao, Peng, and Bai]{GAO2020273}
Tianyu Gao, Hao Peng, and Xiaoli Bai.
\newblock Calibration of atmospheric density model based on gaussian processes.
\newblock \emph{Acta Astronautica}, 168:\penalty0 273--281, 2020.
\newblock \doi{10.1016/j.actaastro.2019.12.014}.

\bibitem[{Boniface, Claude} and {Bruinsma, Sean}(2021)]{refId0}
{Boniface, Claude} and {Bruinsma, Sean}.
\newblock Uncertainty quantification of the dtm2020 thermosphere model.
\newblock \emph{Journal of Space Weather and Space Climate}, 11\penalty0 (53),
  2021.
\newblock \doi{10.1051/swsc/2021034}.

\bibitem[Licata et~al.(2022{\natexlab{b}})Licata, Mehta, Tobiska, and
  Huzurbazar]{licatadendropout}
Richard~J. Licata, Piyush~M. Mehta, W.~Kent Tobiska, and S.~Huzurbazar.
\newblock Machine-learned hasdm thermospheric mass density model with
  uncertainty quantification.
\newblock \emph{Space Weather}, 20\penalty0 (4):\penalty0 e2021SW002915,
  2022{\natexlab{b}}.
\newblock \doi{10.1029/2021SW002915}.

\bibitem[Wilkins and Alfriend(2000)]{wilkins2000characterizing}
Matthew Wilkins and Kyle Alfriend.
\newblock Characterizing orbit uncertainty due to atmospheric uncertainty.
\newblock In \emph{Astrodynamics Specialist Conference}, page 3931, 2000.

\bibitem[Sagnieres and Sharf(2017)]{sagnieres2017uncertainty}
Luc Sagnieres and Inna Sharf.
\newblock Uncertainty characterization of atmospheric density models for orbit
  prediction of space debris.
\newblock In \emph{7th European Conference on Space Debris}, volume~1, pages
  18--21, 2017.

\bibitem[Emmert et~al.(2017)Emmert, Warren, Segerman, Byers, and
  Picone]{EMMERT2017147}
J.T. Emmert, H.P. Warren, A.M. Segerman, J.M. Byers, and J.M. Picone.
\newblock Propagation of atmospheric density errors to satellite orbits.
\newblock \emph{Advances in Space Research}, 59\penalty0 (1):\penalty0
  147--165, 2017.
\newblock ISSN 0273-1177.
\newblock \doi{10.1016/j.asr.2016.07.036}.

\bibitem[Bussy-Virat et~al.(2018)Bussy-Virat, Ridley, and
  Getchius]{BussyVirat18}
Charles~D. Bussy-Virat, Aaron~J. Ridley, and Joel~W. Getchius.
\newblock Effects of uncertainties in the atmospheric density on the
  probability of collision between space objects.
\newblock \emph{Space Weather}, 16\penalty0 (5):\penalty0 519--537, 2018.
\newblock \doi{10.1029/2017SW001705}.

\bibitem[Gondelach et~al.(2022)Gondelach, Linares, and Mun~Siew]{Gondelach22}
David~J. Gondelach, Richard Linares, and Peng Mun~Siew.
\newblock Atmospheric density uncertainty quantification for satellite
  conjunction assessment.
\newblock \emph{Journal of Guidance, Control, and Dynamics}, 2022.
\newblock \doi{10.2514/1.G006481}.

\bibitem[Tobiska et~al.(2021)Tobiska, Bowman, Bouwer, Cruz, Wahl, Pilinski,
  Mehta, and Licata]{sethasdm2021}
W.~Kent Tobiska, Bruce~R. Bowman, S.~David Bouwer, Alfredo Cruz, Kaiya Wahl,
  Marcin~D. Pilinski, Piyush~M. Mehta, and Richard~J. Licata.
\newblock The set hasdm density database.
\newblock \emph{Space Weather}, 19\penalty0 (4):\penalty0 e2020SW002682, 2021.
\newblock \doi{10.1029/2020SW002682}.

\bibitem[Mehta and Linares(2017)]{MehtaL2017SW001642}
Piyush~M. Mehta and Richard Linares.
\newblock A methodology for reduced order modeling and calibration of the upper
  atmosphere.
\newblock \emph{Space Weather}, 15\penalty0 (10):\penalty0 1270--1287, 2017.
\newblock \doi{10.1002/2017SW001642}.

\bibitem[Mehta et~al.(2018)Mehta, Linares, and Sutton]{MLS2018}
Piyush~M. Mehta, Richard Linares, and Eric~K. Sutton.
\newblock A quasi-physical dynamic reduced order model for thermospheric mass
  density via hermitian space-dynamic mode decomposition.
\newblock \emph{Space Weather}, 16\penalty0 (5):\penalty0 569--588, 2018.
\newblock \doi{10.1029/2018SW001840}.

\bibitem[Licata et~al.(2021)Licata, Mehta, Tobiska, Bowman, and
  Pilinski]{QualHASDM}
Richard~J. Licata, Piyush~M. Mehta, W.~Kent Tobiska, Bruce~R. Bowman, and
  Marcin~D. Pilinski.
\newblock {Qualitative and Quantitative Assessment of the SET HASDM Database}.
\newblock \emph{Space Weather}, 19\penalty0 (8), 2021.
\newblock \doi{10.1029/2021SW002798}.

\bibitem[Reigber et~al.(2002)Reigber, Lühr, and Schwintzer]{REIGBER2002129}
Ch. Reigber, H.~Lühr, and P.~Schwintzer.
\newblock Champ mission status.
\newblock \emph{Advances in Space Research}, 30\penalty0 (2):\penalty0
  129--134, 2002.
\newblock ISSN 0273-1177.
\newblock \doi{10.1016/S0273-1177(02)00276-4}.

\bibitem[Sutton(2008)]{suttondiss}
Eric~K. Sutton.
\newblock Effects of solar disturbances on the thermosphere densities and winds
  from champ and grace satellite accelerometer data.
\newblock \emph{Doctoral Dissertation, Department of Aerospace Engineering
  Sciences, University of Colorado, Boulder, Colorado}, 2008.

\bibitem[Mehta et~al.(2017)Mehta, Walker, Sutton, and Godinez]{2016SW001562}
Piyush~M. Mehta, Andrew~C. Walker, Eric~K. Sutton, and Humberto~C. Godinez.
\newblock New density estimates derived using accelerometers on board the champ
  and grace satellites.
\newblock \emph{Space Weather}, 15\penalty0 (4):\penalty0 558--576, 2017.
\newblock \doi{10.1002/2016SW001562}.

\bibitem[Emmert et~al.(2021)Emmert, Drob, Picone, Siskind, Jones~Jr., Mlynczak,
  Bernath, Chu, Doornbos, Funke, Goncharenko, Hervig, Schwartz, Sheese, Vargas,
  Williams, and Yuan]{NRLMSIS2}
J.~T. Emmert, D.~P. Drob, J.~M. Picone, D.~E. Siskind, M.~Jones~Jr., M.~G.
  Mlynczak, P.~F. Bernath, X.~Chu, E.~Doornbos, B.~Funke, L.~P. Goncharenko,
  M.~E. Hervig, M.~J. Schwartz, P.~E. Sheese, F.~Vargas, B.~P. Williams, and
  T.~Yuan.
\newblock Nrlmsis 2.0: A whole-atmosphere empirical model of temperature and
  neutral species densities.
\newblock \emph{Earth and Space Science}, 8\penalty0 (3):\penalty0
  e2020EA001321, 2021.
\newblock \doi{10.1029/2020EA001321}.

\bibitem[Weimer et~al.(2016)Weimer, Sutton, Mlynczak, and Hunt]{intercal}
D.~R. Weimer, E.~K. Sutton, M.~G. Mlynczak, and L.~A. Hunt.
\newblock Intercalibration of neutral density measurements for mapping the
  thermosphere.
\newblock \emph{Journal of Geophysical Research: Space Physics}, 121\penalty0
  (6):\penalty0 5975--5990, 2016.
\newblock \doi{10.1002/2016JA022691}.

\bibitem[Weimer et~al.(2020)Weimer, Mehta, Tobiska, Doornbos, Mlynczak, Drob,
  and Emmert]{extemplar}
D.~R. Weimer, P.~M. Mehta, W.~K. Tobiska, E.~Doornbos, M.~G. Mlynczak, D.~P.
  Drob, and J.~T. Emmert.
\newblock Improving neutral density predictions using exospheric temperatures
  calculated on a geodesic, polyhedral grid.
\newblock \emph{Space Weather}, 18\penalty0 (1):\penalty0 e2019SW002355, 2020.
\newblock \doi{10.1029/2019SW002355}.

\bibitem[Weimer et~al.(2021)Weimer, Tobiska, Mehta, Licata, Drob, and
  Yoshii]{comparison}
Daniel~R. Weimer, W.~Kent Tobiska, Piyush~M. Mehta, R.~J. Licata, Douglas~P.
  Drob, and Jean Yoshii.
\newblock {Comparison of a Neutral Density Model With the SET HASDM Density
  Database}.
\newblock \emph{Space Weather}, 19\penalty0 (12):\penalty0 e2021SW002888, 2021.
\newblock \doi{10.1029/2021SW002888}.

\bibitem[Schutz et~al.(2004)Schutz, Tapley, and Born]{tapley}
Bob Schutz, Byron Tapley, and George Born.
\newblock \emph{Statistical Orbit Determination}, chapter~6.
\newblock Elsevier Academic Press, Burlington, MA, 1 edition, 2004.

\bibitem[McLaughlin et~al.(2012)McLaughlin, Lechtenberg, Fattig, and
  Krishna]{McLaughlin2012}
Craig~A. McLaughlin, Travis Lechtenberg, Eric Fattig, and Dhaval~Mysore
  Krishna.
\newblock Estimating density using precision satellite orbits from multiple
  satellites.
\newblock \emph{The Journal of the Astronautical Sciences}, 59:\penalty0
  84--100, 2012.
\newblock \doi{10.1007/s40295-013-0007-4}.

\bibitem[Kalman(1960)]{10.1115/1.3662552}
R.~E. Kalman.
\newblock {A New Approach to Linear Filtering and Prediction Problems}.
\newblock \emph{Journal of Basic Engineering}, 82\penalty0 (1):\penalty0
  35--45, 03 1960.
\newblock ISSN 0021-9223.
\newblock \doi{10.1115/1.3662552}.

\bibitem[Welch and Bishop(2001)]{Bishop_EKF}
G.~Welch and G.~Bishop.
\newblock {An introduction to the Kalman filter}.
\newblock \emph{University of North Carolina, Department of Computer Science,
  Technical Report TR 95-041}, 2001.

\bibitem[Julier and Uhlmann(1997)]{10.1117/12.280797}
Simon~J. Julier and Jeffrey~K. Uhlmann.
\newblock {New extension of the Kalman filter to nonlinear systems}.
\newblock In \emph{Signal Processing, Sensor Fusion, and Target Recognition
  VI}, volume 3068, pages 182 -- 193. International Society for Optics and
  Photonics, SPIE, 1997.
\newblock \doi{10.1117/12.280797}.

\bibitem[Julier and Uhlmann(2004)]{1271397}
S.J. Julier and J.K. Uhlmann.
\newblock Unscented filtering and nonlinear estimation.
\newblock \emph{Proceedings of the IEEE}, 92\penalty0 (3):\penalty0 401--422,
  2004.
\newblock \doi{10.1109/JPROC.2003.823141}.

\bibitem[Wan and Van Der~Merwe(2000)]{882463}
E.A. Wan and R.~Van Der~Merwe.
\newblock The unscented kalman filter for nonlinear estimation.
\newblock In \emph{Proceedings of the IEEE 2000 Adaptive Systems for Signal
  Processing, Communications, and Control Symposium (Cat. No.00EX373)}, pages
  153--158, 2000.
\newblock \doi{10.1109/ASSPCC.2000.882463}.

\bibitem[Lisano(2006)]{lisano2006nonlinear}
Michael~E Lisano.
\newblock Nonlinear consider covariance analysis using a sigma-point filter
  formulation.
\newblock In \emph{Guidance and Control 2006: Proceedings of the 29th Annual
  AAS Rocky Mountain Guidance and Control Conference}, San Diego, CA, 2006.
  Univelt.

\bibitem[Golub. and Van~Loan(1989)]{golub}
G.H. Golub. and C.F. Van~Loan.
\newblock \emph{Matrix Computations}, chapter~4.
\newblock Johns Hopkins University Press, Baltimore, MD, 2 edition, 1989.

\bibitem[Murray(2018)]{murray2018}
Sophie~A. Murray.
\newblock The importance of ensemble techniques for operational space weather
  forecasting.
\newblock \emph{Space Weather}, 16\penalty0 (7):\penalty0 777--783, 2018.
\newblock \doi{10.1029/2018SW001861}.

\end{thebibliography}






\end{document}